\documentclass[twocolumn,preprintnumbers]{revtex4-1}
\usepackage{amsmath}
\usepackage{latexsym}
\usepackage{graphicx}
\usepackage{slashed}
\usepackage{pifont}
\usepackage{color}
\usepackage[usenames,dvipsnames]{xcolor}

\newcommand{\req}[1]{Eq.\,(\ref{#1})}
\newcommand{\beqn}{\begin{equation}}
\newcommand{\eeqn}{\end{equation}}

\newcommand{\TCP}{T_{\rm CEP}}

\newcommand{\CEP}{CEP\:}

\begin{document}

\preprint{MIT-CTP/4783}

\title{Universal relations between nongaussian fluctuations in heavy-ion collisions}
\author{Jiunn-Wei Chen$^{1,2,*}$, Jian Deng$^{3,\dag}$, Hiroaki Kohyama$^{1,\flat}$ and Lance Labun$^{4,\sharp}$}
\affiliation{$^{1}$ Department of Physics, CTS and LeCosPA,
National Taiwan University, Taipei, 10617 Taiwan\\
$^2$ Center for Theoretical Physics, Massachusetts Institute of Technology, Cambridge, MA 02139, USA \\
$^3$ Key Laboratory of Particle Physics and Particle Irradiation (MOE),
School of Physics, Shandong University, Jinan 250100, China\\
$^4$ Department of Physics, The University of Texas, Austin, Texas 78746 USA}
\email{jwc@phys.ntu.edu.tw \\ $^\dag$ jdeng@sdu.edu.cn \\ $^\flat$ kohyama.hiroaki@gmail.com \\ $^\sharp$ labun@utexas.edu}
\date{14 March, 2016}

\begin{abstract}
We show that universality near a critical end point implies a characteristic relation between third- and fourth-order baryon susceptibilities $\chi_3$ and $\chi_4$, resulting in a banana-shaped loop when $\chi_4$ is plotted as a function of $\chi_3$ along a freeze-out line.  This result relies only on the derivative relation between $\chi_3$ and $\chi_4$, the enhancement of the correlation length and the scaling symmetry near a critical point, and the freeze out line near the critical point not too parallel to the $\mu_B$ axis.  Including the individual enhancements of $\chi_3$ and $\chi_4$ near a critical point, these features may be a consistent set of observations supporting the interpretation of baryon fluctuations data as arising from criticality.
\end{abstract}

\maketitle



{\bf Introduction}---
Determining the existence of a critical point in the QCD phase diagram would provide invaluable information on the behaviour of nuclear matter at high density, indicating the existence of a first order phase transition line extending to higher density and lower temperature \cite{QCDphases}.  Models of chiral symmetry breaking suggest a critical end point (\CEP) may exist at high enough temperature and low enough density to be accessible to heavy-ion collisions at small center-of-mass energy per nucleon $\sqrt{s_{NN}}$, motivating the beam energy scan at RHIC \cite{Mohanty:2009vb,Aggarwal:2010cw,Gupta:2011wh,Heinz:2015tua,Luo:2015doi} and future work at FAIR.  Predicting signals of a critical point in QCD is made challenging by strong-coupling in analytic theory and the sign-problem on the lattice.  The motivated hypothesis of a critical point allows us to use the universality of critical phenomena to study the impact of a \CEP on some observables.  

One universal characteristic of a critical point is diverging correlation length $\xi\to\infty$, due to the order parameter field becoming massless.  Higher order fluctuation moments of observables coupled to the order parameter also diverge, in particular baryon number and transverse momentum \cite{Stephanov:2008qz,Stephanov:2011pb}.  These susceptibilities can be measured by event-by-event fluctuations in heavy-ion collision experiments \cite{Stephanov:1998dy}.  The approach via universality requires assuming the existence of a critical point. It yields robust qualitative features  which are supported by models calculations near the \CEP \cite{Asakawa:2009aj,Fu:2010ay,Skokov:2011rq,Chen:2014ufa,Chen:2015dra}. However, it remains a challenge to predict quantitatively the magnitude of a signal in heavy ion collisions.

We show that a diverging correlation length implies a strict ordering of features in the third- and fourth-order baryon number susceptibilities, which offers a robust qualitative signature of criticality.
Baryon number is conserved in QCD reactions. The baryon susceptibilities are calculated as the derivatives of the logarithm of the partition function $\chi_n=
-\partial^n \ln \mathcal{Z}/\partial\mu_B^n$ and are measured as event-by-event fluctuations in heavy ion collisions \cite{Aggarwal:2010wy,Adamczyk:2013dal}.  The third-order (skewness) and fourth-order (kurtosis)
\begin{align}\label{chi3chi4}
\chi_3
=\frac{\partial\chi_2}{\partial\mu_B},
\quad
\chi_4=\frac{\partial\chi_3}{\partial\mu_B}
\end{align}
are expected to provide stronger signals of criticality because they diverge with a larger power of $\xi$ than the second order susceptibility \cite{Stephanov:2008qz}.
As indicated, they are related to each other and the second-order susceptibility by a $\mu_B$ derivative.  For comparison to experimental observables, volume dependence is eliminated in the ratios
\begin{align}
m_1=\frac{T\chi_3}{\chi_2}
\qquad
m_2=\frac{T^2\chi_4}{\chi_2}\,.
\end{align}
In the vicinity of a critical point, these ratios exhibit divergences and non-monotonic behaviour and differ from the analogous ratios of net proton number susceptibilities only by non-singular contributions, which should be sub-leading \cite{Hatta:2003wn}.  Therefore, in the scaling region the behaviour of $m_1,m_2$(baryon) should predict qualitatively the behaviour of net proton susceptibilies $m_1,m_2$(proton), which are measured in experiment.

Using the 3D Ising model which is in the same universality class as QCD, we show that: (a) Universality implies that, along a freeze-out line passing near a critical end point, third and fourth-order fluctuation moments are strictly ordered 
\begin{align}\label{Tordering}
T_{{\rm min},m_2}>T_{{\rm max},m_1}>T_{{\rm max}, m_2}>\TCP\,.
\end{align}
(b) Obtaining signatures of criticality does not require the experiment achieving freeze-out at temperature equal or lower than $\TCP$.  Scaling may be visible at significantly higher $T$, similar to the hypothesized quantum critical point in the phase diagram of high-$T_c$ superconductors which is at $T=0$ and masked by a superconducting regime \cite{highTcSC}. The magnitudes of $m_1$ and $m_2$ may provide a handle on the distance from the critical point. This will be demonstrated in the NJL model with deformed couplings.

Taken together with measurements of $\chi_3$ and $\chi_4$ which are expected to exhibit independently particular divergences and functional dependences, this correlation of $\chi_3$ and $\chi_4$ along a freeze-out line is a consistency check that would support interpretation of the experimental data as bona fide signatures of criticality.  Our discussion focusses on the equilibrium susceptibilities, but the real system created in heavy-ion collisions is an expanding gas that may or may not be in local equilibrium.  In particular, long wavelength modes are subject to critical slowing, with the result that the expansion generally reduces the correlation length from its equilibrium expectation value \cite{slowing}.  This slowing of the correlation length induces a memory effect: the correlation length and higher order susceptibilities reflect the state of the system at an earlier time in the expansion \cite{Mukherjee:2015swa,Mukherjee:2016kyu}.  Notably, even with the memory effect, the relative distribution of positive and negative regions of kurtosis (Fig. 3 of \cite{Mukherjee:2015swa}) appears to preserve its equilibrium relation to the skewness, \req{chi3chi4}, and it will be interesting to see how our equilibrium results are modified by the memory effect. 

{\bf Interplay between Skewness and Kurtosis}---As we will show in the next section, the qualitative relation between $\chi_3$ and $\chi_4$ in QCD only relies on (a) the derivative relation between $\chi_n$ in \req{chi3chi4}, and (b) the enhanced correlation length and corresponding scaling symmetry near a CEP. Here we start by studying order parameter fluctuations in the 3D Ising model, which belongs to the same universality class as QCD, just to illustrate our points. Our analysis does not rely on the mapping between Ising model and QCD. However, we will also see that if the CEP is too close to the $T$ axis, such that the phase boundary and freeze-out line near the CEP parallel the $\mu_B$ axis, the signature banana shape of the $m_1$-$m_2$ plot becomes degenerate and can even change qualitatively.


In the Ising model, the general coordinates are $(H,t)$ with $H$ the external magnetic field and $t \equiv (T-T_c)/T_c$ the reduced temperature.  The order parameter, the magnetization $M=-\partial \ln \mathcal{Z}/\partial H$, is an odd function of $H$. $M$ is discontinuous at $H = 0$ for $t<0$.  The \CEP is at the origin where the correlation length $\xi$ diverges. The non-analytic (long-distance) behavior near the \CEP is universal among systems of the same universality class while the analytic (short-distance) behavior is model dependent. The non-analytic part of the equation of state $M=M(H,t)$ near the \CEP has the scaling symmetry: if $M$ is known for a specific positive\,(negative) $t$, then $M$ for all positive\,(negative) $t$ is known. 

\begin{figure}
\includegraphics[width=0.45\textwidth]{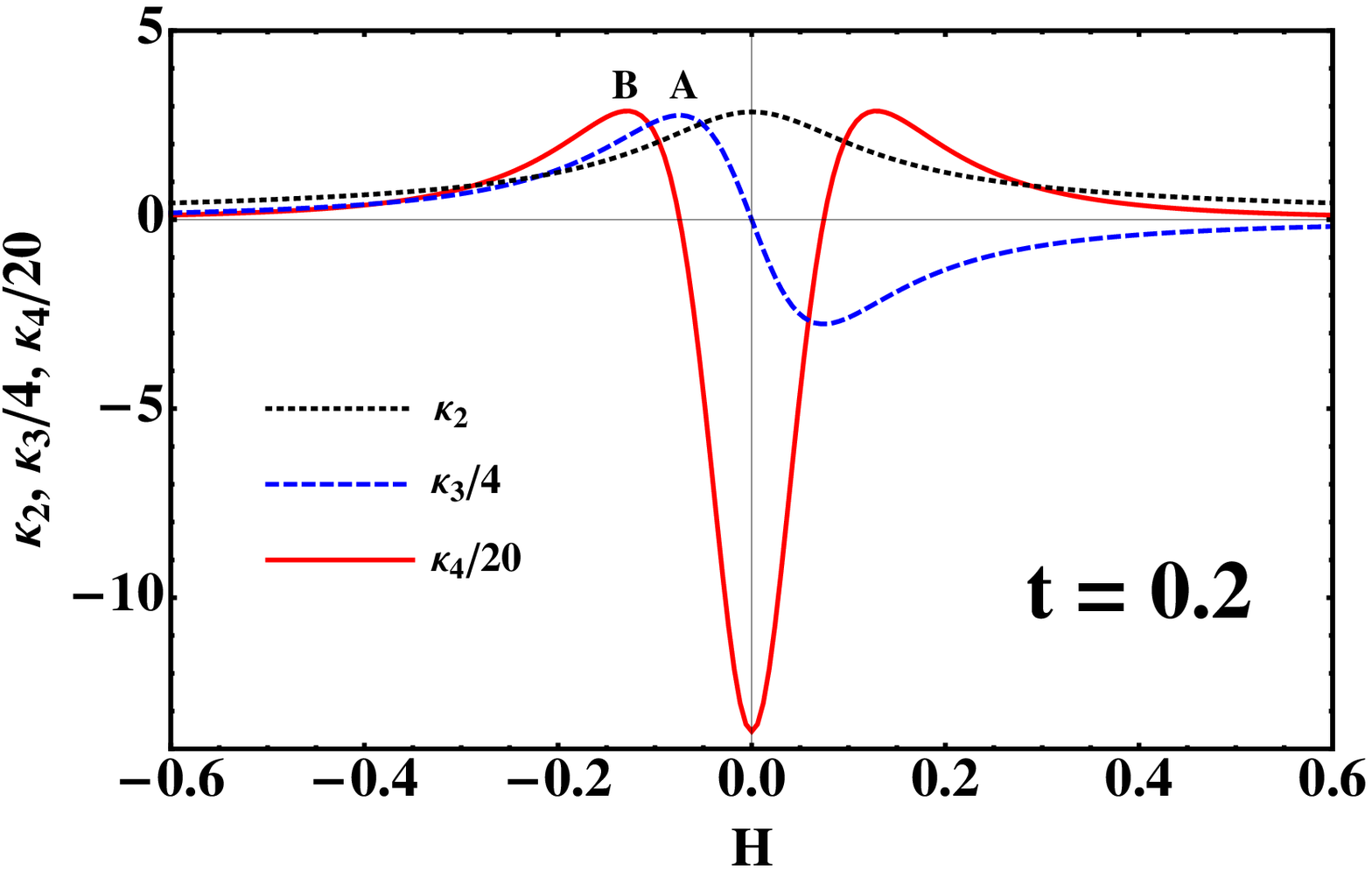}
\includegraphics[width=0.45\textwidth]{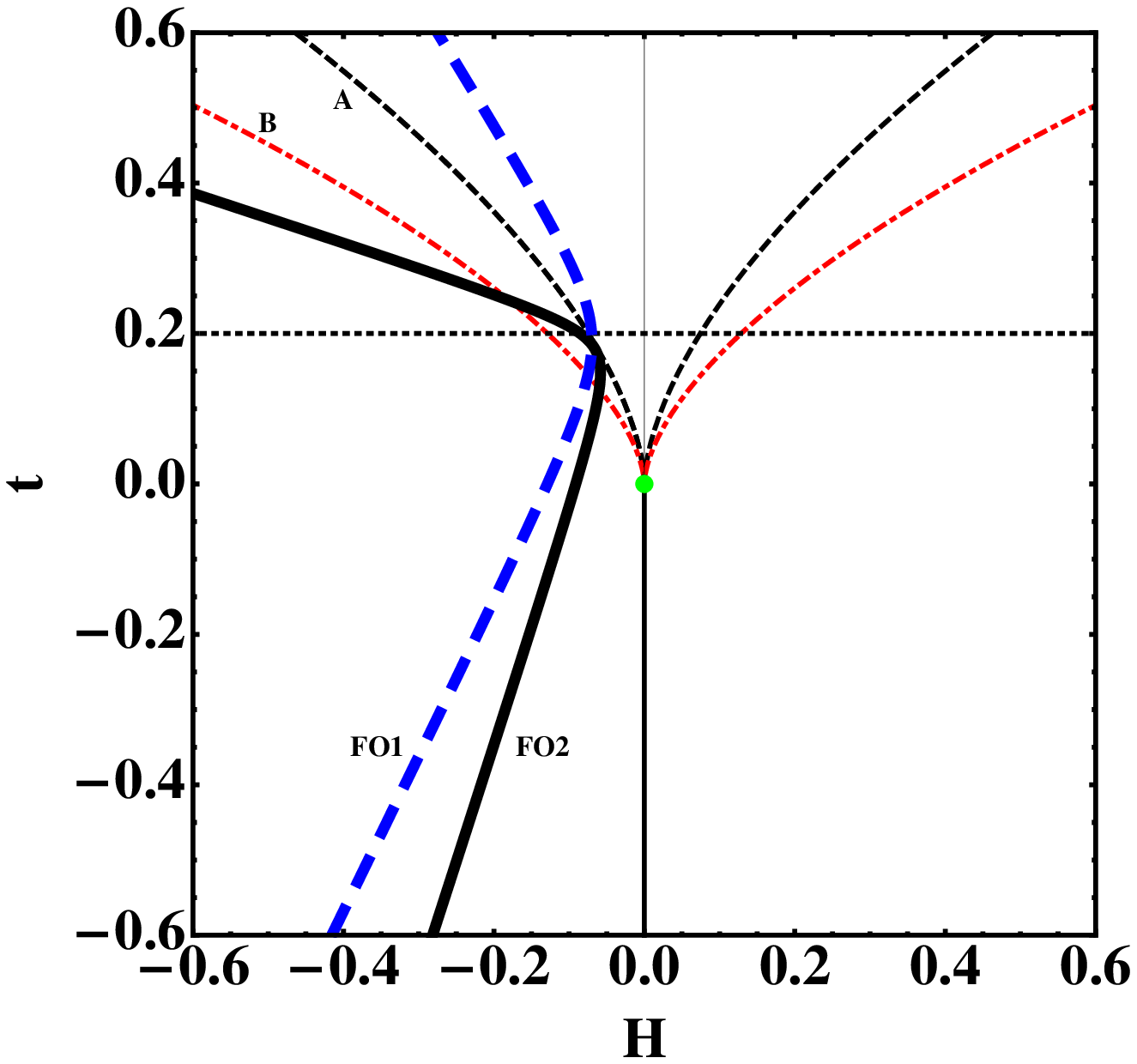}
\caption{Upper panel: $\protect\kappa_{2,3,4}(H)$ at fixed $t>0$.  Lower panel: the Ising model phase diagram with Line A the maximum  of $\kappa_3$ (also $\kappa_4=0$), and Line B the maximum of $\kappa_4$.  The curved lines are example freeze-out lines, drawn to model how they may pass through the scaling region in QCD.}
\label{fig:kappaH}
\end{figure}

The susceptibilities are derivatives with respect to $H$, 
\begin{align}\label{kappa3kappa4}
\kappa_2\equiv \frac{\partial M}{\partial H},
\quad
\kappa_3
=\frac{\partial\kappa_2}{\partial H}\,,
\quad
\kappa_4=\frac{\partial\kappa_3}{\partial H},
\end{align}
paralleling the relation of  $\chi_{n+1}$ to $\chi_n$, each $\kappa_{n+1}$ is related to $\kappa_n$ by one derivative with respect to the classical source. We first concentrate on susceptibilities at $t=0.2$, obtaining susceptibilities at all $t>0$ by the scaling symmetry mentioned above. We expect that 
$\kappa_2 \propto\xi^2$ is a decreasing function of $|H|$ since $\xi$ is larger when closer to the \CEP. Also, $\kappa_2$ is a smooth function of $H$ since $t=0.2$ is in the crossover region. Therefore, $\kappa_2 $ should have a smooth peak centered at $H=0$ and then asymptote to zero as $H \rightarrow \infty$ as shown in the upper panel of Fig. \ref{fig:kappaH}, which is computed using the non-analytic part of the equation of state of Ref. \cite{IsingEOS}.

\begin{figure}[t]
\includegraphics[width=0.22\textwidth]{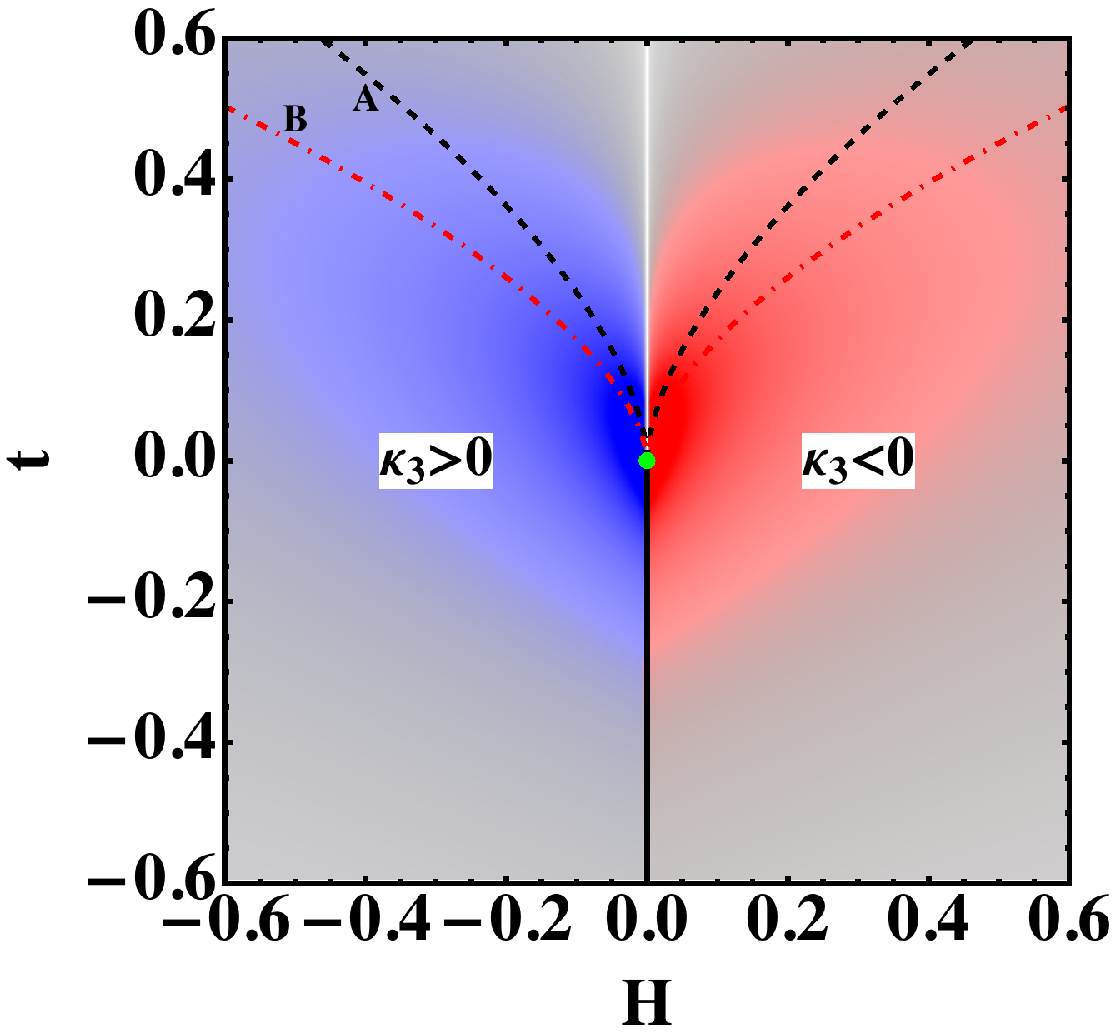}
\includegraphics[width=0.22\textwidth]{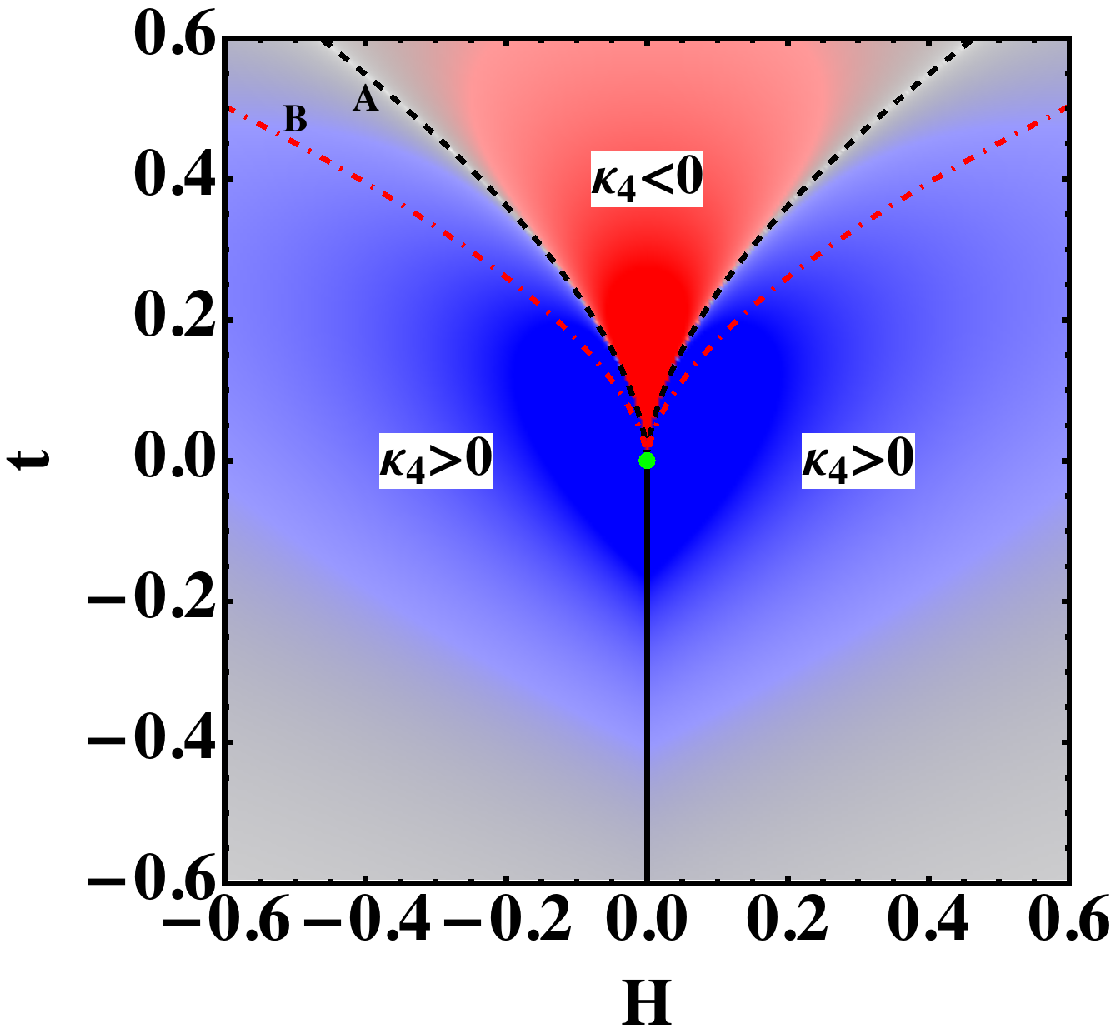}\\
\includegraphics[width=0.38\textwidth]{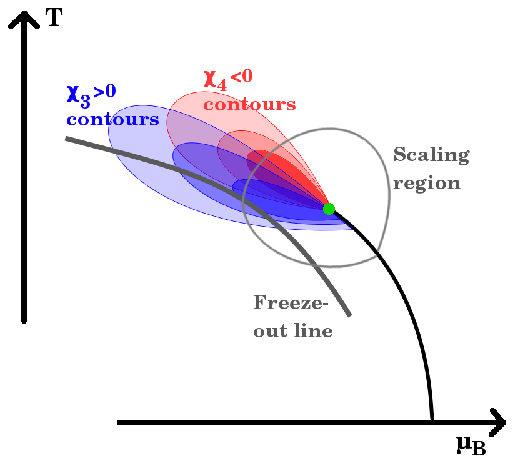}
\caption{ Upper left (right): density plot of $\kappa_3\,(\kappa_4)$ in the Ising model.  Regions of $\kappa_i>0$ are in blue and $\kappa_i<0$ are in red.  The dotted (black) line is the same as Line A in Fig.\,\ref{fig:kappaH} and dot-dashed (red) line the same as Line B.  Lower panel:  A sketch of the peaks in $\chi_3$ and $\chi_4$ on a plausible phase diagram of QCD together with a hypothetical freeze-out line.  Comparison to the location of the maxima in $\chi_3$ and $\chi_4$ in Fig.~\ref{fig:kappaH} suggests how the freeze-out line may be mapped into the Ising coordinates.  }
\label{fig:sketch}
\end{figure}

Once $\kappa_2$ is obtained, its derivatives yield $\kappa_3$ and $\kappa_4$, also shown in the upper panel of Fig. \ref{fig:kappaH}. It is easy to see that, for $H<0$, the peak of $\kappa_3$ coincides with $\kappa_4 = 0$ while the peak of $\kappa_4$ is at larger $|H|$ than the peak of $\kappa_3$. Then by scaling symmetry, the peaks of  $\kappa_3$  and  $\kappa_4$ lie on lines A and B, respectively, in the lower panel of Fig. \ref{fig:kappaH}. At constant $t<0$, however, $\kappa_{2-4}$ are decreasing functions of $|H|$ with no inflection points. The density plots of $\kappa_{3(4)}$  is shown in the upper panel of Fig.~\ref{fig:sketch} with a $H$-odd(even) heart-shaped pattern.

In QCD, the story is very similar. We expect the most singular, non-analytic behavior near the \CEP has scaling symmetry as well. For a constant $T > T_{\CEP}$, $\chi_2$ has the same single peak behavior as $\kappa_2$ in the Ising model at $t = 0.2$ shown in Fig. \ref{fig:kappaH}. Then $\chi_3$ and $\chi_4$ are obtained by taking derivatives of $\chi_2$, similar to analysis of the Ising model. Therefore, we expect $\chi_3$ and $\chi_4$ contours in the scaling region in QCD behave as depicted in the lower panel of Fig.~\ref{fig:sketch} analogous to what happens in the Ising model. This feature is robust and is confirmed in model calculations \cite{Asakawa:2009aj,Fu:2010ay,Skokov:2011rq,Chen:2014ufa,Chen:2015dra}.  We will show that the anti-clockwise loop in the $\kappa_3$-$\kappa_4$ plane corresponds to an anti-clockwise loop in the $m_1$-$m_2$ plane.  Moreover, the loop and its direction only depend on the relation between $m_1$ and $m_2$ (paralleling that between $\kappa_3$ and $\kappa_4$) and the existence of a critical end point at $\mu_B>0$, whether or not the \CEP is accessible to experiment.

\begin{figure}[t]
\includegraphics[width=0.4\textwidth]{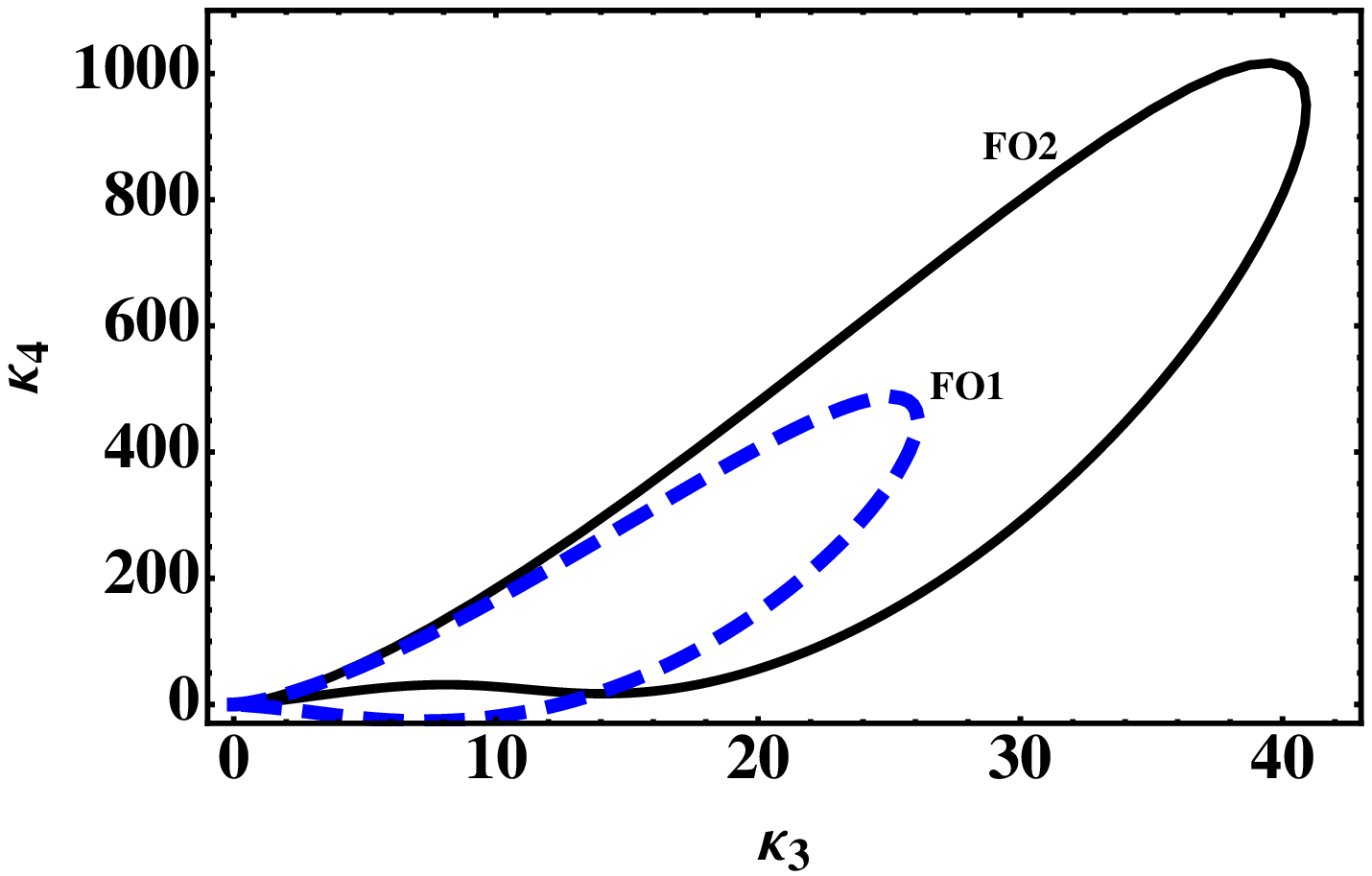}
\caption{$\kappa_4$ versus $\kappa_3$ on example freeze-out lines passing through the universal region as shown in Fig.~\ref{fig:kappaH}.  Temperature decreases in the anti-clockwise direction.}
\label{fig:banana}
\end{figure}

{\bf Scenarios in Energy Scan Experiments}---In experiments, one does not know whether the \CEP exists and even if it does, what its location is \emph{a priori}. Therefore, the question is, what can one learn from the susceptibilities measured on the freeze out line? 

To answer this question, we consider the following scenarios for the phase boundary between the hadronic phase and the quark gluon plasma phase: (1) a cross over at higher $T$ and a first order at lower $T$ connected by a \CEP (this is the typically assumed scenario), (2) purely cross over, or (3) purely first order. (2) and (3) can be considered as generalizations of (1), with the \CEP for (2) located in the ``fourth quadrant" of the phase diagram ($T<0,\mu_B>0$) while the \CEP of (3) located in the ``second quadrant" ($T>0,\mu_B<0$).



{\bf \emph{ Scenario I: \CEP at $T >0$}}---In this scenario, contours of $\chi_3$ and $\chi_4$ near the \CEP are as depicted in the lower panel of Fig.~\ref{fig:sketch}. The location of the \CEP is not known precisely and neither are the contours.  The boundary of $\chi_4 = 0$ is determined by the cancellation of the leading singular structure and can be shifted by subleading, model-dependent, analytic contributions. Our strategy is to draw a few generic freeze out lines, depicted in the lower panels of Fig.~\ref{fig:sketch} and Fig. \ref{fig:kappaH}, then ask whether there are common features of susceptibilities on those lines. In Fig. \ref{fig:kappaH},  we assume that the freeze out line is a function of $t$.  Going from high to low $t$, the simplest case is FO1, which crosses lines A and B once each. The corresponding $\kappa_4$-$\kappa_3$ curve is shown in Fig. \ref{fig:banana} with the curve going anti-clockwise forming a ``banana'' shape from high to low $t$. This figure shows the ordering
\begin{align}\label{kappaordering}
t_{{\rm min},\kappa_4}>t_{{\rm max},\kappa_3}>t_{{\rm max},\kappa_4}>0\,,
\end{align}
necessarily arises from the derivative relation between the $\kappa_n$ and $\kappa_{n+1}$.  All features occur at temperature higher than the critical point temperature. As the fluctuations become larger closer to the \CEP, the closer the freeze out line to the \CEP, the larger and more elongated the banana is.

In Fig. \ref{fig:kappaH}, we also consider a freeze out line FO2 that crosses line B twice. The corresponding $\kappa_4$-$\kappa_3$ plot in Fig. \ref{fig:banana} also has the banana shape but has two local maximum peaks in $\kappa_4$. Those features remain when one plots $m_2$-$m_1$ instead of $\kappa_4$-$\kappa_3$ since $\kappa_2$ changes slowly when $\kappa_{3(4)}$ changes rapidly. 

One can draw other possible freeze out lines, but the feature of an anti-clock wise loop remains, provided the line remains in the $H<0$ half-plane as is physically sensible for freeze-out in the hadronic phase. This can be seen from the fact that at high $t$, the freeze out line can start from the regime above line A, between lines A and B, or below line B, while at low $t$, it goes below line B. This implies these freeze out lines at high and low $t$ will look very similar to FO1 and FO2 in Fig. \ref{fig:banana} near the origin. This is enough to decide the loop is anti-clock wise which is a feature in common with experiment data \cite{Aggarwal:2010wy,Adamczyk:2013dal,Luo:2015ewa}.

{\bf \emph{Scenario II: \CEP at $T\lesssim 0$}}---As we argue above, the banana shape in $m_2$-$m_1$ is due to the scaling symmetry governed by the \CEP. But could this connection be so strong such that the banana shape is observable even if the \CEP is at $T=0$ or even $T < 0$? One example is high-$T_c$ superconductors \cite{highTcSC}. It is hypothesized that there is a quantum critical point at $T\simeq 0$ that controls the scaling symmetry at finite $T$ but is masked by a superconducting regime. If this is also the case in QCD, then seeing the banana shape in $m_2$-$m_1$ might just suggest the existence of scaling symmetry, which is likely due to a critical point, even while the critical point is at $T < 0$ and hence technically not on the phase diagram.  

\begin{figure}[!t]
\begin{picture}(300,150)
\put(0,0){\includegraphics[width=0.48\textwidth]{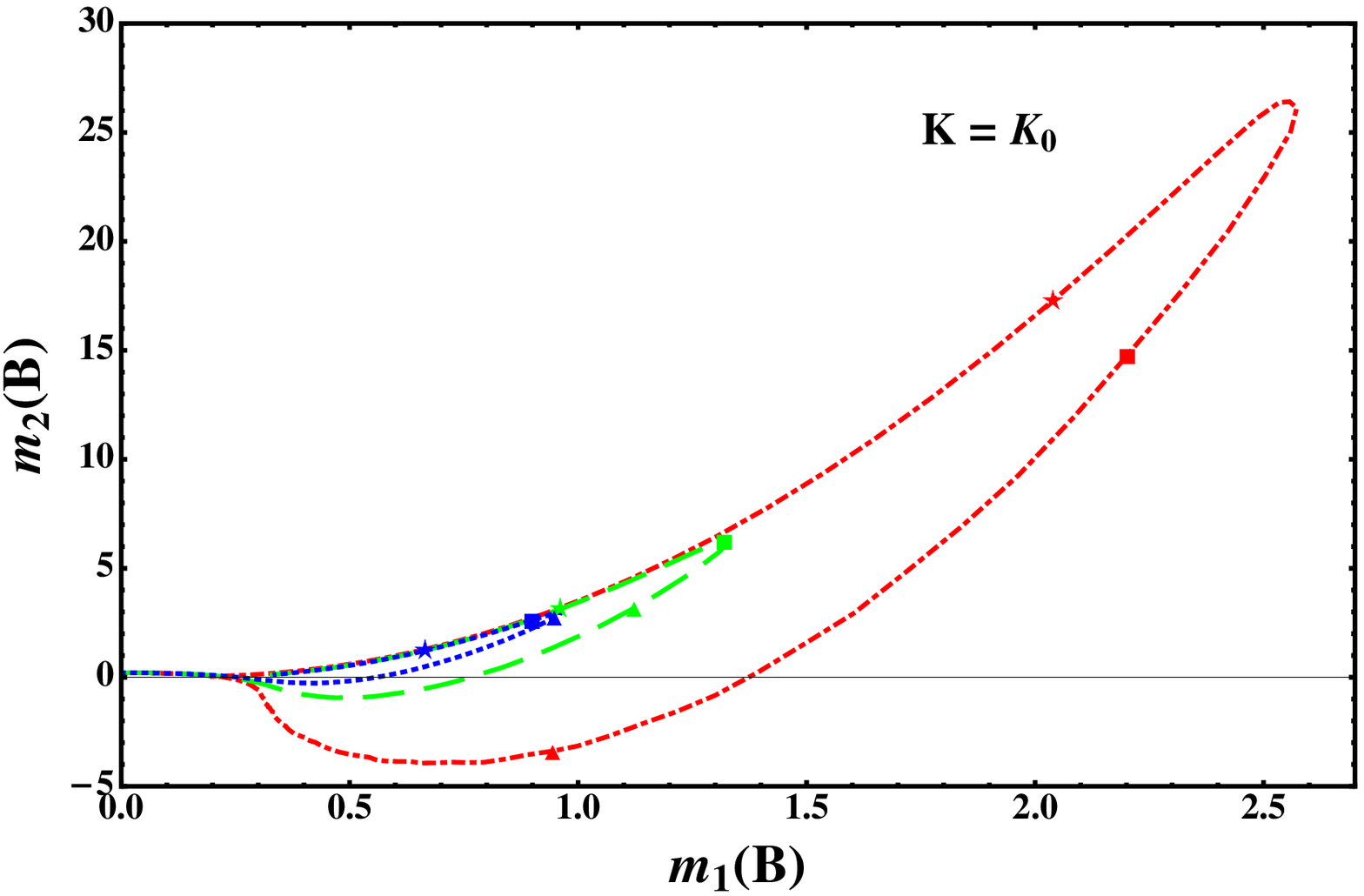}}
\put(30,75){\includegraphics[width=0.24\textwidth]{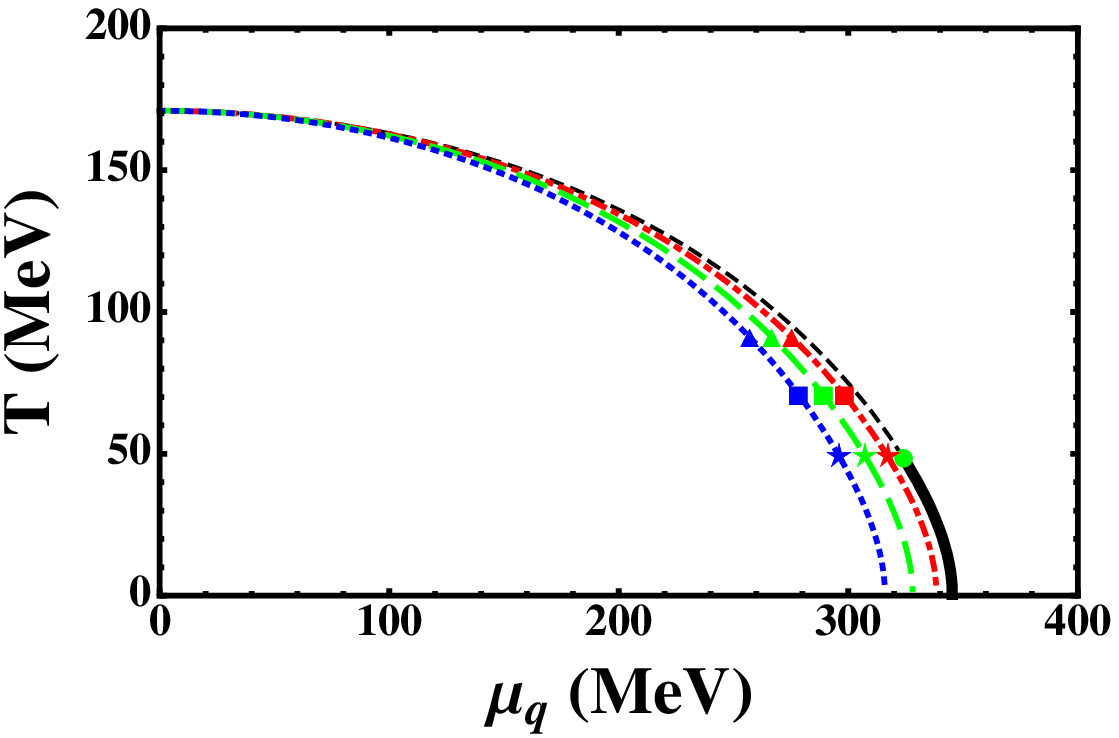}}
\end{picture}
\caption{Inset: the phase diagram of the NJL model with $K=K_0$, a value chosen to reproduce QCD observables, and three hypothetical freeze-out lines tracking the phase boundary (see text).  Larger frame: $m_2$ versus $m_1$ on the freeze-out lines plotted in the inset. }
\label{fig:K1}
\end{figure}

\begin{figure}[!t]
\begin{picture}(300,150)
\put(0,0){\includegraphics[width=0.48\textwidth]{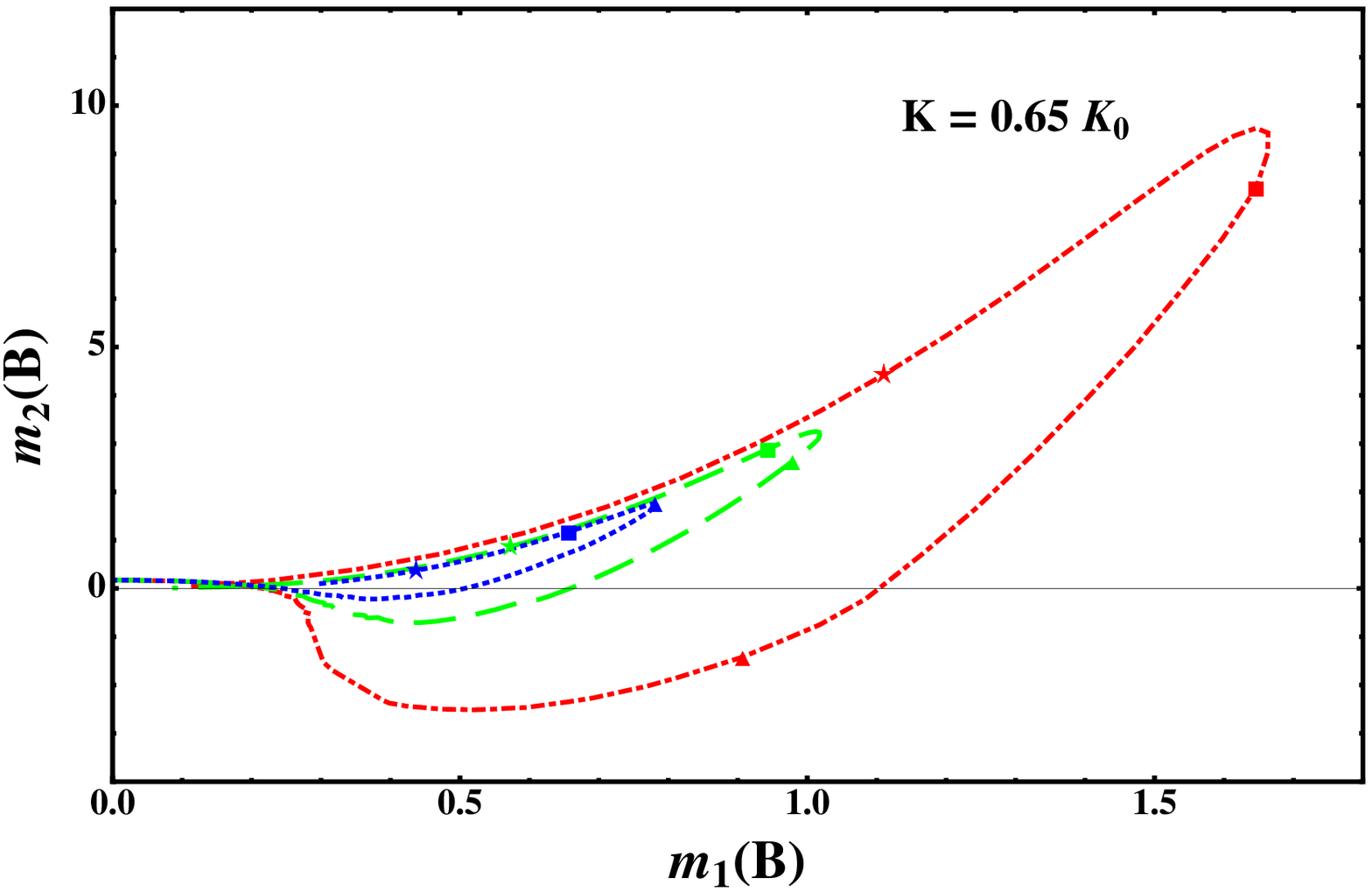}}
\put(25,75){\includegraphics[width=0.24\textwidth]{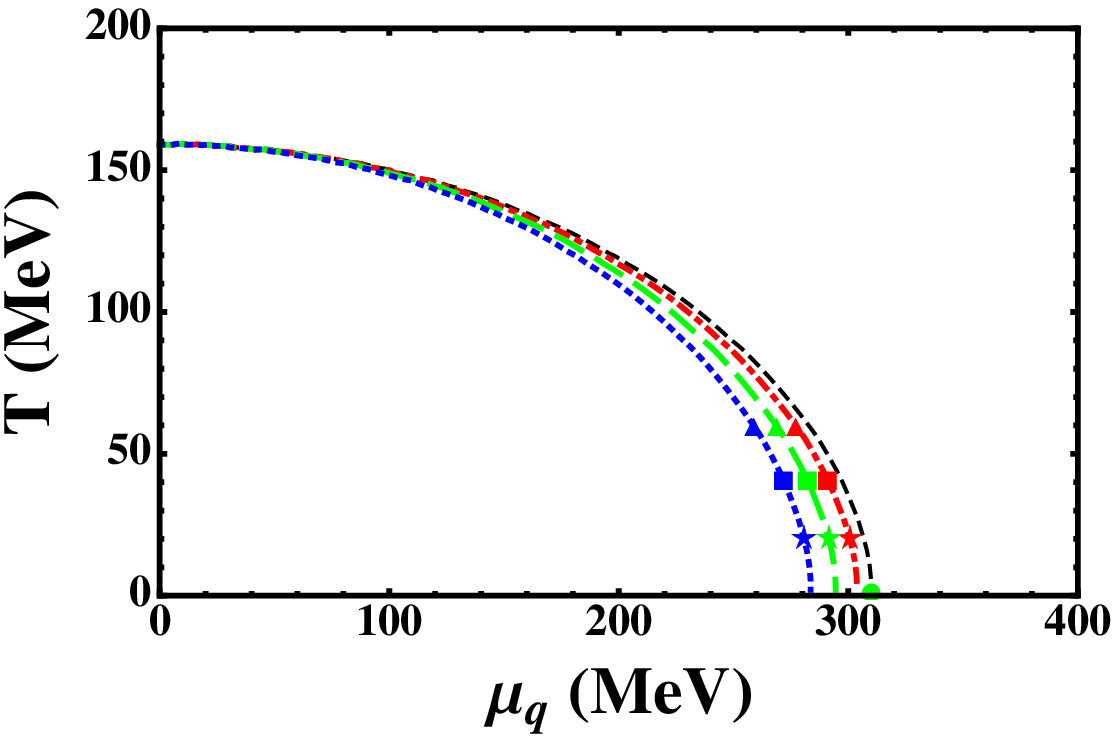}}
\end{picture}
\caption{ Same as Fig.\,\ref{fig:K1} but with $K=0.65K_0$.  The critical end point is at $\TCP=0$.}  
\label{fig:K065}
\end{figure}

\begin{figure}[!th]
\begin{picture}(300,170)
\put(0,0){\includegraphics[width=0.48\textwidth]{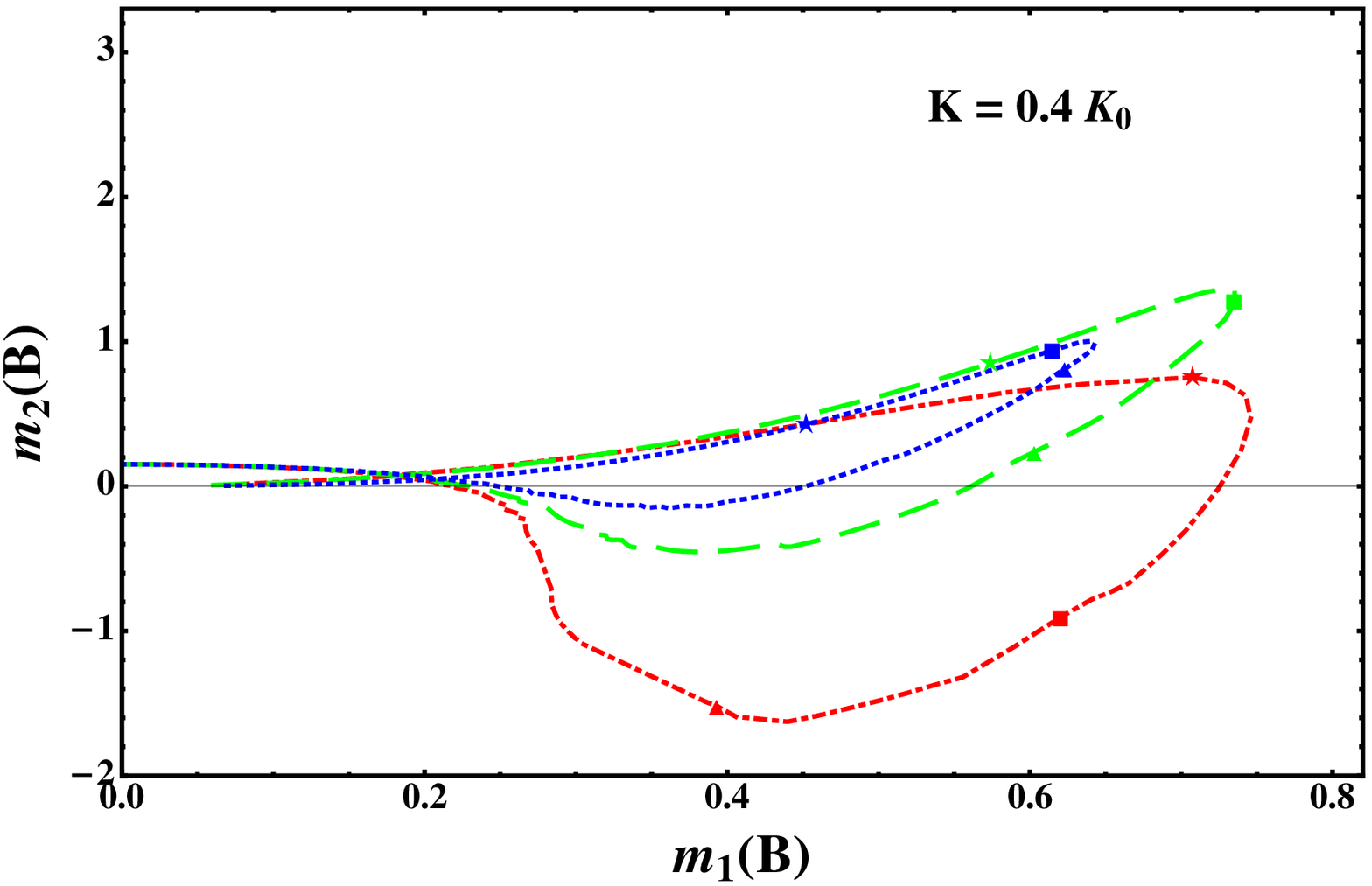}}
\put(25,80){\includegraphics[width=0.22\textwidth]{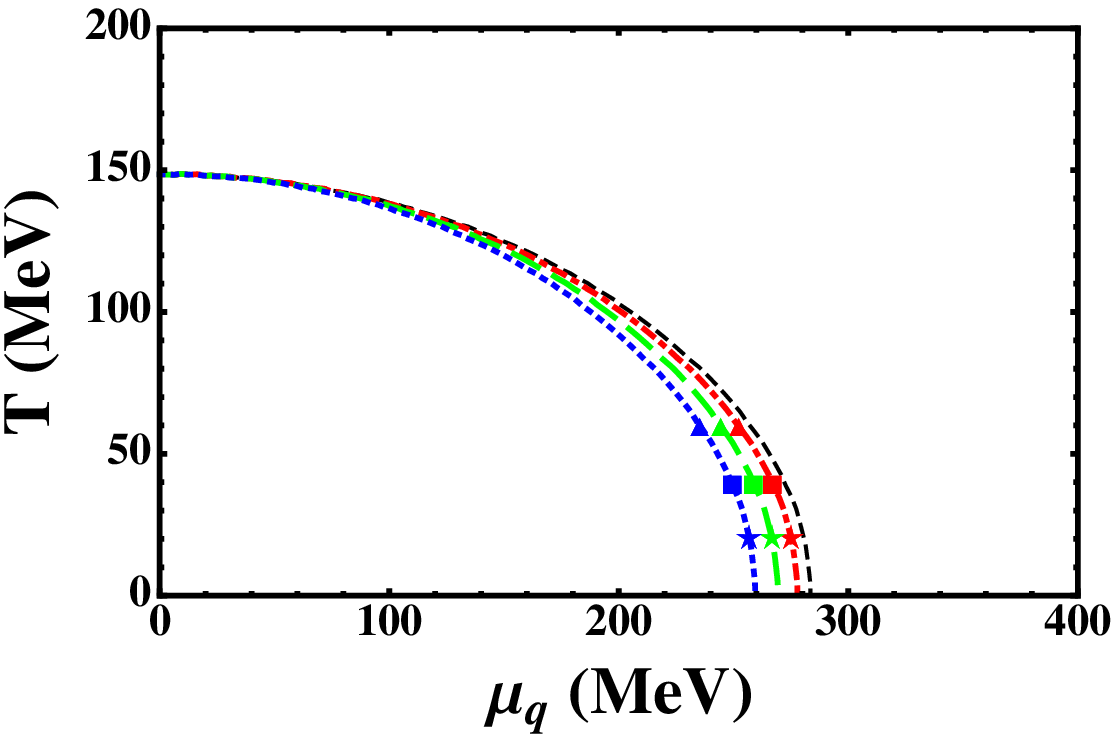}}
\end{picture}
\caption{Same as Figure \ref{fig:K065} with $K=0.4K_0$.  The critical end point would be formally at $\TCP<0$.  Note the difference in scale of the axes. }
\label{fig:K04}
\end{figure}

To study this interesting scenario, we vary the $K$ parameter that controls the anomaly-induced 6-fermion interaction in the Nambu--Jona-Lasinio (NJL) model \cite{NJL} away from the most-used value 
$K_0$ to reduce $\TCP$.  We find that when $K=0.65 K_0$, $\TCP \simeq 0$ and when $K=0.4 K_0$, $\TCP$ effectively becomes negative by extrapolation. The phase boundaries for $K=1.0,0.65,0.4 K_0$ are shown in the insets in Figures \ref{fig:K1}, \ref{fig:K065} and \ref{fig:K04}, respectively, together with the hypothetical freeze out lines which are obtained by rescaling $\mu_B$ of the phase boundaries by factors of $0.98,0.95,0.915$ for the red, green, blue curves, respectively. We have also plotted the corresponding $m_2$-$m_1$ plots in Figs. \ref{fig:K1}-\ref{fig:K04} which exhibit the following features: 

(1) The anti-clockwise banana behavior could survive even $\TCP \le 0$.
If freeze-out can occur at or below $\TCP$, all three features are visible: the minimum in $m_2$, maximum in $m_1$ and maximum in $m_2$.  However, if $\TCP$ is too low to be reached by the experimental conditions, only a subset of these features appears in the data.  Which subset provides a rough guide to how far away $\TCP$ is, as well as an estimation of the upper limit of $\TCP$.

(2) The magnitude of the susceptibilities also changes significantly with the distance from a critical point. This is because the dimensionless $m_{1(2)}$ scales with positive powers of $\xi$ divided by typical scales in the system set by $\mu_B$ and $T$, and $m_2$ scales with more powers of $\xi$ than $m_1$. Therefore when the freeze out line is far away from the \CEP, both $m_1$ and $m_2$ are limited to small values $\lesssim 1$ and $m_2$ is the same magnitude as $m_1$. 
Nearer a critical point, $m_2$ achieves values significantly greater than 1. This is the reason $m_1$-$m_2$ banana plots become larger and more elongated for freeze-out lines in closer proximity to a critical point.


{\bf Scenario III: \CEP formally at $\mu_B<0$}--In the limit of very small quark masses, the QCD phase transition is first order in the physical region $\mu_B,T>0$.  Although excluded by experiment, this scenario is interesting because it helps reveal the effect of the mapping from QCD variables to universal Ising coordinates $(\mu_B,T)\to (H,t)$.  In this limit, the phase boundary typically becomes increasingly parallel to the $\mu_B$ axis as $\mu_B\to 0$.  Approaching this limit smoothly from having the \CEP at $\mu_B,T>0$ (Scenario I), we can infer how the $m_1$-$m_2$ plot changes as the phase boundary becomes more parallel to the $\mu_B$ axis even while the \CEP remains in the first quadrant.

\begin{figure}[!t]
\begin{picture}(300,150)
\put(0,0){\includegraphics[width=0.48\textwidth]{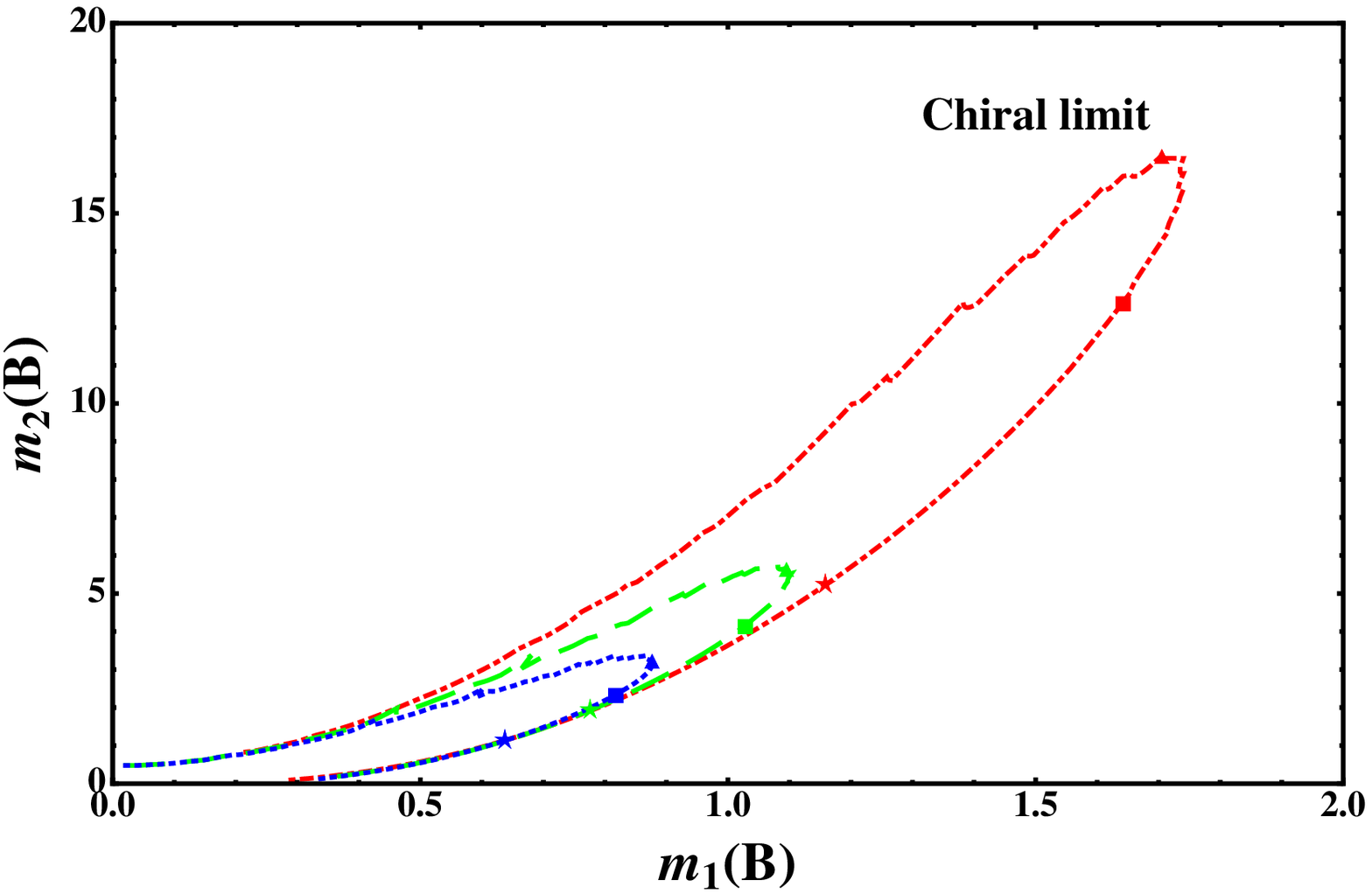}}
\put(25,75){\includegraphics[width=0.23\textwidth]{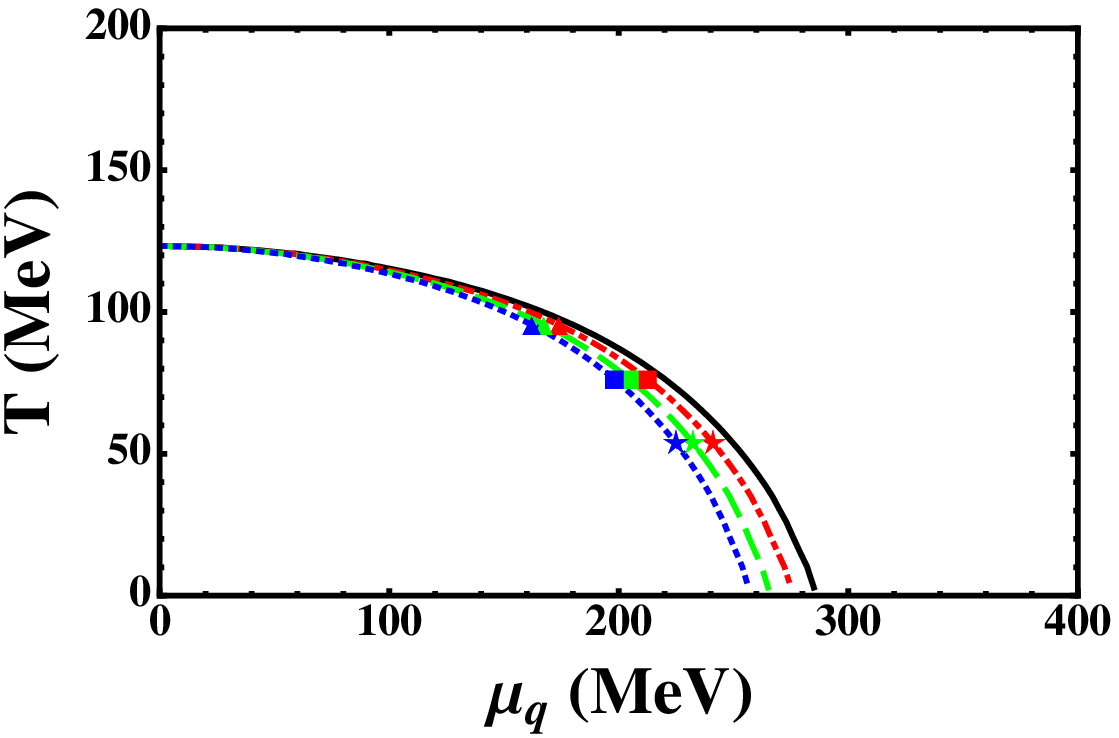}}
\end{picture}
\caption{ Same as Fig.\,\ref{fig:K1} but with all quark masses zero $m_u=m_d=m_s=0$. The phase boundary is the first order line. There is no critical point in the first quadrant $\mu_B,T>0$, or the critical end point can be considered in the second quadrant with $\mu_{CEP}<0$.}  
\label{fig:chiral}
\end{figure}

The NJL model calculation of the $m_2$-$m_1$ plot in this limit is shown in Fig.\,\ref{fig:chiral}. We see a clock-wise behavior.  This can be understood by analyzing the effective potential, or equivalently the log of the partition function, near $\mu_B=0$.  In this region, a Taylor expansion $\ln\mathcal{Z}=z_0+z_2\mu_B^2/2+z_4\mu_B^4/4!+...$ involves only even powers of $\mu_B$ due to particle-anti-particle symmetry.  This implies that $m_1\simeq (z_4/z_2) \mu_B$ while $m_2\propto (z_4/z_2)$ as $\mu_B\to 0$. We expect the negative $m_2$ region exists only in the second quadrant, therefore,  $z_4/z_2>0$, and the freeze-out curve in the $m_2$-$m_1$ plane starts at $m_1\to 0,m_2>0$ at the high collision energy/low $\mu_B$ end.  At the other end of the curve, we consider the $T\to 0$ limit, where fluctuations are increasingly Gaussian due to the increasing distance from the phase boundary and the dominance of the quadratic term near the bottom of the potential.  Thus, at low collision energy/high $\mu_B$ end of the curve, we expect $m_1,m_2\to 0$.  Since $m_1\geq 0$ everywhere in between, the curve connecting these two limits must follow a \emph{clockwise} trajectory in the $m_1$-$m_2$ plane, in stark contrast to the anti-clockwise loop seen above.

Seeing that the loop changes from anti-clockwise to clockwise as $\mu_{CEP}\to 0$, the loop must become increasingly narrow, passing through the degenerate case in which the loop collapses to a line or crossese itself, in order to change its orientation between the two limits.  This agrees with our numerical experiments in the NJL model, which we are able to move from Scenario I toward Scenario III by tuning the bare quark mass to very small values.  More systematic exploration of the quark mass dependence of the observables may be taken up in future publications.

\begin{figure}[t]
\includegraphics[width=0.48\textwidth]{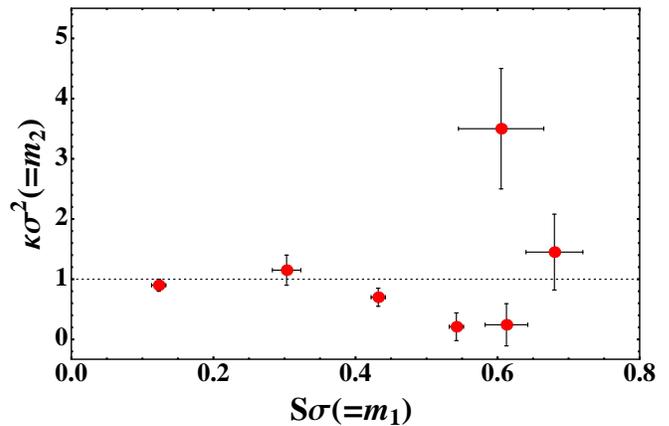}
\caption{From STAR preliminary data for the net-proton distribution in 0-5$\%$ centrality of Au$+$Au collisions\,\cite{Luo:2015ewa}, $\kappa \sigma^2$ versus $S \sigma$ forms an anticlockwise loop from high to low $\sqrt{s_{NN}}=200,62.4,39,27,19.6,11.5,7.7$ GeV. }
\label{fig:data-banana}
\end{figure}

{\bf Summary}---Putting these observations together, we may formulate rough criteria for evidence of proximity to a critical point.  If $m_2$ achieves maximum $\gtrsim 2$, accompanied by $m_1$ maximum $\gtrsim 1$, and the maxima obey the ordering \req{Tordering}, then freeze-out is occurring near enough to a critical point for universality to provide the leading order dynamics.  

The important feature of scaling in determining this behaviour is the non-monotonic behaviour of the correlation length along a freeze-out line.  Different hypothetical freeze-out lines are best discussed in Ising coordinates: for $\mu_B$ increasing ($\sqrt{s_{NN}}$ decreasing) the freeze-out line must proceed from high $t\gg 0$ to low $t<0$.  Then, whether the freeze-out line is taken at constant $H$ as in Ref.\,\cite{Stephanov:2011pb} or curved as in Figure \ref{fig:kappaH}, the correlation length $\xi^2\sim\kappa_2$ has a maximum along the line.  Moreover, by hypothesis this maximum is the sought after signature of singular behaviour, implying that the skewness and kurtosis are also enhanced and the ratios $m_1,m_2$ should be $\gg 1$.  This enhancement should be robust in presence of other, model-dependent dynamics.  On the other hand, without a critical point and associated scaling, the enhancements of the correlation length and higher moments vanish, and the features predicted above are expected to be no stronger than thermal fluctuations and other dynamics of the system.

Preliminary data for $m_1$ and $m_2$ \cite{Luo:2015ewa} shown in Figure \ref{fig:data-banana} suggest that the lowest collision energies may be entering the scaling region, seeing that $m_2/m_1\sim 4\gg 1$, but not reaching as low as $\TCP$, seeing that the banana shape is not complete.  With the present data, we cannot yet know whether the lowest energy point at $\sqrt{s_{NN}}=7.7$ GeV belongs to the upper branch of the banana seen in the preceding figures.  Acquiring data in the range $7.7 \leq \sqrt{s_{NN}}\leq 11.5$ GeV could help determine the relative location of a critical point by clarifying the shape and magnitude of the banana and comparing to the different scenarios depicted in Figs. \ref{fig:K1}-\ref{fig:K04}.

There are of course experimental challenges associated with measuring $\chi_3$ and $\chi_4$, such as effects from dynamics of the fireball expansion \cite{slowing} and beam energy-dependent final-state hadron-gas interactions.  These effects must be understood and accounted for in the data as well as possible to enhance any signal of criticality.  Our purpose has been to argue that, if freeze-out occurs near enough to a critical point (in the scaling region) for nongaussian fluctuations to be impacted as predicted in preceding work, then the qualitative relation between the higher-order fluctuation moments \req{Tordering} could hold.  This can serve as a first check that the experimentally observed peaks in $m_1$ and $m_2$ arise from an underlying divergence of the correlation length in the fireball.  We anticipate that several more such consistent sets of characteristics should be found in order to prove that the experimental data do reflect the presence of a critical point.


\vskip0.2cm 
\textit{Acknowledgments}:  J.D. is supported in part by the Major State Basic Research Development Program in China (Contract No. 2014CB845406), National Natural Science Foundation of China (Projects No. 11105082).  J.-W.C. is supported in part by the MOST and the NTU-CASTS of R.O.C.  H.K. is supported by Ministry of Science and Technology (Taiwan, ROC), through Grant No. MOST 103-2811-M-002-087.   L.L. is supported by NNSA cooperative agreement de-na0002008, the Defense Advanced Research Projects Agencys PULSE program (12-63-PULSE-FP014), the Air Force Office of Scientific Research (FA9550-14-1-0045) and the National Institute of Health SBIR 1 LPT\_001.


\end{document}